\documentclass[preprint,5p,times]{elsarticle}

\usepackage[T1]{fontenc}
\usepackage{float}
\usepackage{amsmath}
\usepackage{amsthm}
\usepackage{bm}
\usepackage{breakurl}
\usepackage{breqn}
\usepackage{amssymb}
\usepackage{array} 
\usepackage{varwidth} 
\usepackage{listings}
\usepackage{gensymb}
\usepackage{fancyhdr}
\usepackage{url}
\usepackage{lineno}
\usepackage{xcolor,colortbl}
\usepackage{multirow}
\usepackage{algorithm}
\usepackage{algpseudocode}
\usepackage{mathtools}
\usepackage{graphicx}

\usepackage{svg}
\usepackage{csvsimple}
\usepackage{makecell}
\usepackage{rotating}
\usepackage{nomencl}
\usepackage{subcaption}

\usepackage[english]{babel}

\usepackage{amsmath}
\usepackage{graphicx}
\usepackage[colorlinks=true, allcolors=blue]{hyperref}

\author[a]{Enzo Meneses\corref{author}}
\author[a]{Crist\'obal A. Navarro}
\author[a]{H\'ector Ferrada}
\author[b]{Konstantin Verichev}
\author[c]{Cristian Salazar-Concha}

\cortext[author] {Corresponding author.\\\textit{E-mail address:} enzo.meneses@alumnos.uacl.cl}
\address[a]{Instituto de Informática, Universidad Austral de Chile.}
\address[b]{Instituto de Obras Civiles, Universidad Austral de Chile.}
\address[c]{Instituto de Administración, Universidad Austral de Chile.}

\makenomenclature

\journal{arXiv}

\begin{document}
\begin{frontmatter}

\title{Ray Tracing Cores for General-Purpose Computing: A Literature Review}


\begin{abstract}
Ray tracing cores are specialized units in modern GPUs designed to accelerate real-time ray tracing, enhancing rendering in gaming, animation, and visualization. Beyond graphics, recent research has explored repurposing these cores to solve non-graphical problems by reformulating them as geometric queries, leveraging the inherent parallelism of ray tracing. Although successful in specific cases, these applications lack a clear pattern, and the conditions under which RT cores can provide computational benefits are still not clearly understood. The objective of this literature review is to examine diverse applications of ray tracing cores in general-purpose computation, identifying common features, performance gains, and limitations. By categorizing these efforts, the review aims to provide guidance on the types of problems that can effectively exploit ray tracing hardware beyond traditional rendering tasks. This is achieved with a blibliometric review based on 59 research articles indexed in Scopus, and a systematic literature review on 35 of them which propose new RT solutions and compare them with state-of-the-art methods to solve 32 distinct problems, in some works achieving up to $200\times$ speedup. Most of the problems analyzed in this work have applications in physics simulations and in solving some geometric queries, but problems with potential applications in databases and AI can also be found. Analyzing the characteristics of the problems, it was found that nearest neighbor search, including its variants, benefit the most from ray tracing cores as well as problems that rely on heuristic to diminish the necessary work. This is aligned with the biggest strength of RT cores; discarding tree branches when traversing a tree to avoid unnecessary work. Also, it was found that many short-length rays should be preferred over a few large rays. The results found in this work can serve as a guide for knowing beforehand which applications are better potential candidates to benefit from RT Core computation.

\end{abstract}


\end{frontmatter}

    \nomenclature{ANN}{Approximate Nearest Neighbor}
    \nomenclature{kNN}{k-nearest Neighbor}
    \nomenclature{FRNN}{Fixed-radius Nearest Neighbor}
    \nomenclature{CD}{Collision Detection}
    \nomenclature{RMQ}{Range Minimum Query}
    \nomenclature{SpMM}{Sparse Matrix Multiplication}
    \nomenclature{DBSCAN}{Density-Based Spatial Clustering of Applications with Noise}
    \nomenclature{FEM}{Finite Element Method}
    \nomenclature{SM}{Streaming Multiprocessor}
    \nomenclature{BVH}{Bounding Volume Hierarchy}
    \nomenclature{HPC}{High Performance Computing}

\printnomenclature

\section{Introduction}

High-performance computing (HPC) has become indispensable for advancing scientific discovery and engineering, particularly in domains that require massive data processing and complex simulations~\cite{Dongarra2024TheCO,Koch2023HPCIT}. 
In artificial intelligence, training large-scale deep learning models demands billions of operations per second and substantial memory bandwidth. The largest models consist of trillions of parameters~\cite{Fedus2021SwitchTS} and thousands of GPUs are used in their training~\cite{Narayanan2021EfficientLL}. 
Similarly, projects such as AbacusSummit~\cite{Maksimova2021AbacusSummitAM} demonstrate the power of HPC for physics: the project produced a massive set of high-accuracy, high-resolution N-body simulations to study cosmic structure formation using 60 trillion bodies. 
In microbiology, a molecular dynamics simulation directly contributed to the research of the COVID-19 vaccine with simulations of 83 million atoms~\cite{Schade2022BreakingTE}.
These examples highlight the growing demand for compute power and its importance in scientific development.

Meeting these computational demands comes at a considerable resource cost.
Large HPC centers consume enormous amounts of electricity and require sophisticated cooling systems to maintain operational stability~\cite{CocanaFernandez2019EcoEfficientRM}.
Scaling systems to meet current demand involves managing millions of cores, high-bandwidth interconnects, and massive memory hierarchies, as illustrated by a case study of NERSC Perlmutter system~\cite{Li2023AnalyzingRU}.
Furthermore, resource requirements extend beyond the sheer core counts; large-scale workloads, such as those of AI or astrophysical simulations, require substantial storage, memory, and I/O bandwidth, demonstrating that HPC resource planning must encompass the entire computational ecosystem~\cite{Gerber2011LargeSC}.
Due to the difficulty in meeting these demands, future HPC systems are expected to be over-provisioned and power-constrained, requiring a careful balance between performance and energy consumption~\cite{Arima2024OnTC}. Furthermore, with Dennard and frequency scaling no longer providing automatic efficiency gains, software performance optimization becomes an increasingly important factor to solve bigger problems~\cite{Martin2014PostDennardSA,Sueur2010DynamicVA}.

While computational demands require multiple resources, and there are several approaches to reduce consumption, HPC focus lies on improving time performance. Recent studies have shown that reducing time-to-solution is the first step in improving energy efficiency~\cite{Suarez2025EnergyET}. Thus, with traditional HPC techniques and properly taking advantage of current hardware it is possible to solve more/bigger problems in a reasonable time. In this context, graphics processing units (GPUs) have emerged as a key enabler of high-throughput parallel computation in modern HPC systems. Originally designed for graphic tasks, GPUs evolved into general-purpose programmable accelerators capable of executing a wide range of computationally intensive tasks~\cite{owens2008gpu,Dally2021EvolutionOT}. Their massive parallelism, supported by thousands of cores and high-bandwidth memory, allows HPC applications to achieve unprecedented performance~\cite{navarro2014survey}. However, GPUs are optimized for uniform, data-parallel workloads and encounter performance limitations when handling hierarchical or irregular data structures. For instance, ray tracing~\cite{Aila2009UnderstandingTE}, sparse matrices~\cite{Boukaram2019HierarchicalMO}, and other tree-based computations exhibit divergent control flow and non-contiguous memory access patterns that reduce GPU efficiency. These challenges underscore the need for algorithmic redesigns or specialized hardware to fully leverage GPU capabilities. Moreover, the end of Moore's law further motivates it~\cite{Theis2017TheEO,Dally2020DomainspecificHA}.

\begin{figure*}[ht!]
  \centering
  \includegraphics[scale=0.45]{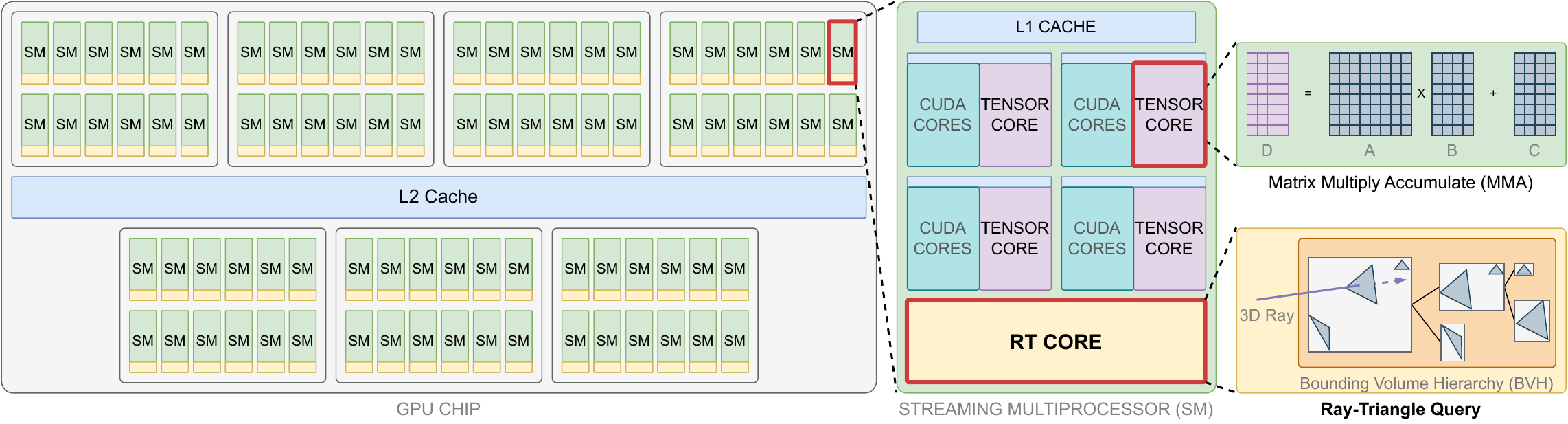}
  \caption{Modern GPU architecture design.}
  \label{fig:gpu-arch}
\end{figure*}

In order to keep up with the increasing compute demands, modern GPUs have incorporated application-specific integrated circuits (ASICs), specifically, tensor cores and RT cores. Figure~\ref{fig:gpu-arch} shows the architecture design of current GPUs. As an example, the NVIDIA RTX 5090 with Blackwell architecture has 21,760 CUDA cores, 680 tensor cores and 170 RT cores among 170 SMs~\cite{NVIDIA_Blackwell_whitepaper}. Tensor Cores enable high-throughput matrix operations optimized for deep learning and scientific computing, dramatically improving performance for AI-driven simulations~\cite{Markidis2018NVIDIATC}.
Similarly, linear algebra operations experience comparable improvements~\cite{Lopez2023MixedPL}.
Further research also showed that other computation patterns can be represented as matrix multiplications and take advantage of tensor cores, for example, cellular automata~\cite{Navarro2024CATCA} or prefix-sum~\cite{dakkak2019accelerating}.

Regarding ray tracing cores (RT cores), they were introduced to allow real-time ray tracing in video games and accelerate film animation rendering~\cite{Sanzharov2020SurveyON,Meister2021ASO}. Ray tracing is used primarily in visualization tasks~\cite{Sarton2023StateoftheartIL}. Also in computer vision, including robotics and autonomous driving, by creating a visualization of a 3D model and comparing it with the images taken by the robot~\cite{Echeverra2024HarnessingTP}. RT cores mitigate some of the inefficiencies associated with irregular data structures in the GPU. However, since ray tracing is such an specific problem, RT cores often remain underutilized, or not utilized at all, in general purpose compute tasks.
Therefore, further research has been performed to extend the use of RT cores by reformulating other problems as ray tracing queries, for example, nearest neighbor search~\cite{zhu2022rtnn}, range queries~\cite{Meneses2024RTXRMQ} and database indexing~\cite{Henneberg2023RTIndeXEH}, among many others.

In the case of neighbor search, more specifically fixed-radius nearest neighbor, the key insight is to reverse the way the query is constructed. Normally, a sphere is built around the query point to find all the points inside the sphere. On the other hand, the ray tracing approach builds a sphere for each point of the problem, then launches a ray on the coordinates of the query points and finds all the spheres that were hit by the ray. Other variants of nearest neighbor are built upon this, for example, TrueKNN iteratively increases the search radius while performing fixed radius search until the k-nearest neighbors are found~\cite{Nagarajan2023RTkNNSUU}. This has also been used as part of a solution to other types of problems such as clustering~\cite{Nagarajan2023RTDBSCANAD} and particle simulations~\cite{zhao2023leveraging}.

Research on expanding the versatility of RT cores presents two main challenges: i) reformulating a problem into the ray tracing model and ii) identifying which problems will indeed benefit from this RT Core reformulation. The underlying conditions that allow a problem to be more efficient with RT cores are not fully understood yet.
Literature reviews play an essential role in evaluating breadth, depth and future possibilities of a particular research area~\cite{Grant2009ATO}. It also reinforces the categorization of the research and analysis of studies in areas related to the field~\cite{Chigbu2023TheSO}. Systematic literature reviews (SLR) are particularly valuable due to the quality of the results they produce, improving accuracy and information dissemination~\cite{Budgen2006PerformingSL,Liberati2009ThePS}. A  bibliometric analysis measures the development and trends in a particular field, evaluating scientific progress and its evolution, and highlighting research gaps for further exploration~\cite{Borner2012NetworkST}. 
The objective of this literature review is to evaluate the use of RT cores for general-purpose computing, identifying common features and limitations, as well as their impact on performance. For this purpose, a bibliometric analysis is used to understand trends about the fields and problems in which RT cores have been employed. Then, a systematic literature review is used to analyze the specific problems solved with ray tracing and their performance compared to state-of-the-art methods.

Five research questions are addressed throughout the analysis. (\textbf{RQ1}) What are the current bibliometric trends in using RT cores for general-purpose computation? (\textbf{RQ2}) What are the non-graphic applications of RT cores, and how are computational problems reformulated as RT queries? (\textbf{RQ3}) How much does performance improve when using RT cores? (\textbf{RQ4}) What key features are frequently present in the problems that best adapt to RT cores? (\textbf{RQ5}) What are common limitations in the use of RT cores for general-purpose computing?

The subsequent sections of this work are structured as follows. Section 2 presents the methodology applied in the study, followed by the results of the literature review in section 3. Section 4 presents a discussion of the results. And finally, section 5 offers the conclusion of the study.

\section{Methodology}
\begin{figure*}[ht!]
  \centering
  \includegraphics[width=\textwidth]{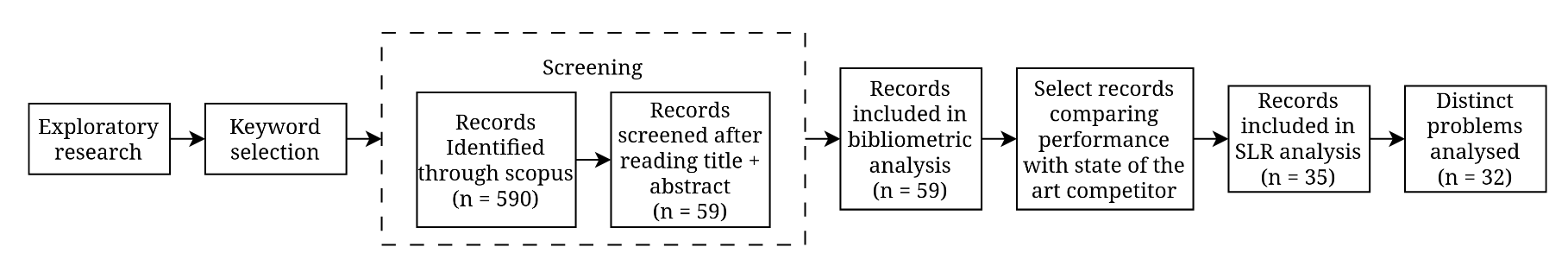}
  \caption{Methodology used}
  \label{fig:methodology}
\end{figure*}

The methodology used is presented in Figure~\ref{fig:methodology}, which is inspired by some PRISMA~\cite{Page2020TheP2} guidelines.
The exploratory research consists of a preliminary search of scientific publications using RT cores beyond graphics, focused on identifying keywords, synonyms and abbreviations used in the field.
With these keywords the search parameters are defined, iteratively looking for keywords previously missed and focusing on increasing the breadth of the obtained articles. The final search query is shown in Table~\ref{table:query}. The first part of the query includes keywords related to GPUs and NVIDIA. AMD is included for completeness but no research on ray tracing was found on their GPUs. This is consistent with the current scientific HPC ecosystem, where AMD's adoption is less than NVIDIA due to their limited tooling. In addition, AMD took longer to offer hardware-accelerated ray tracing and its performance is inferior. The second part includes keywords related to RT cores and different ways to refer to them. Ray-traced was also included to consider articles prior to RTX GPUs with innovative uses of ray tracing. The research query was applied to all fields of documents from Scopus, which indexes relevant conference and Web of Science (WOS) articles.

\begin{table*}[]
\centering
\caption{Search query and criteria used to select documents for this study}
\begin{tabular}{|ll|}
\hline
\multicolumn{1}{|l|}{GPU related keywords}      & (gpu? OR gpgpu OR cuda OR nvidia* OR rtx OR amd* OR rocm)                                                                                                                                                                                                                                                                 \\ \hline
\multicolumn{2}{|c|}{AND}                                                                                                                                                                                                                                                                                                                                           \\ \hline
\multicolumn{1}{|l|}{RT cores related keywords} & \begin{tabular}[c]{@{}l@{}}("ray tracing core?" OR "rt core?"OR "hardware accelerated ray tracing"\\ OR  "ray traced" OR "ray tracing hardware" OR "hardware ray tracing"\\ OR "ray tracing unit?" OR raytraced  OR "raytracing core?"\\  OR "ray tracing accelerator?" OR "ray tracing technology")\end{tabular} \\ \hline
\multicolumn{1}{|l|}{Inclusion criteria}        & \begin{tabular}[c]{@{}l@{}}- Use of ray tracing beyond graphics\\  - Reformulation of a problem to the ray tracing model\end{tabular}                                                                                                                                                                             \\ \hline
\multicolumn{1}{|l|}{Exclusion criteria}        & \begin{tabular}[c]{@{}l@{}}- Graphic tasks\\ - Computer vision tasks\end{tabular}                                                                                                                                                                                                                                 \\ \hline
\multicolumn{1}{|l|}{Date}                      & 06-Oct-2025                                                                                                                                                                                                                                                                                                       \\ \hline
\end{tabular}
\label{table:query}
\end{table*}

Once the search query yielded the general results, 590 documents, they were manually filtered according to the criteria shown in Table~\ref{table:query}. This process was realized by one person reading the title, abstract and keywords of the documents, and in some cases skimming the full document. The selected articles consist of research on the use of RT cores outside of graphics or the application of these solutions to other problems. This excludes all articles related to visualization and computer vision, unless it is about a specific step in the process or unrelated to ray tracing, like space skipping and segmentation respectively.

With the 59 documents selected from these filters a bibliometric analysis using biblioshiny~\cite{Aria2017bibliometrixAR} is performed to obtain an understanding of the problems and areas where RT cores are used. After reading the papers, those including a performance comparison between ray tracing approaches and state-of-the-art methods were selected for the systematic literature review. This resulted in 35 articles covering 32 distinct non-graphics problems. To analyze the selected documents and the use of ray tracing, this work proposes categorizing the problems by different taxonomies. The aim is to find categories that group problems showing similar behavior in terms of the improvement produced by the use of ray tracing relative to the state-of-the-art solution. Then, based on these categories, the analysis tries to explain the reason for that behavior and expand the understanding to take better advantage of RT cores.

\section{Results and Discussion}

\begin{figure}[ht!]
  \centering
  \includegraphics[width=0.45\textwidth]{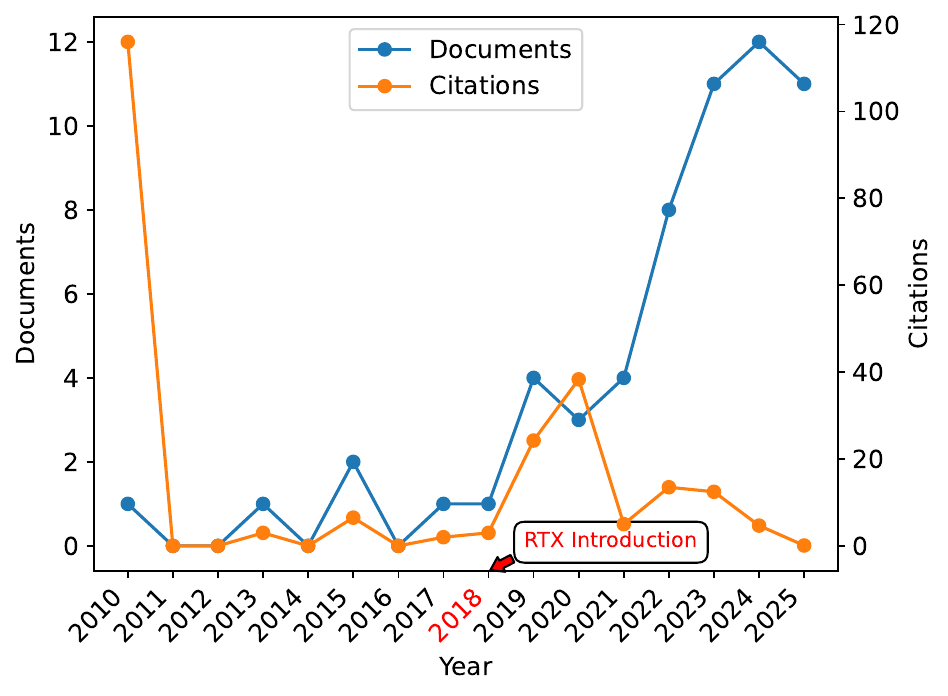}
  \caption{Annual production and average citations per document per year}
  \label{fig:annual_production}
\end{figure}

The main  information of the 59 documents used for the bibliometric analysis is presented in Table~\ref{table:main_information}. The publications are authored by 173 authors with an average of almost 4 authors per document. Only three are by a single author. They were published in 49 sources with an annual growth rate of 17.33\%. The documents are evenly divided into articles and conference papers. There is also one review analyzing iso-contouring methods~\cite{Buscher2024ACS}, with ray tracing being one the methods discussed. Figure~\ref{fig:annual_production} shows the annual publication and annual citation starting in 2010. Before the introduction of RT cores in November 2018, the production was consistently below two publications per year. Only with the advantage of hardware acceleration did the research start to grow. Regarding the annual citation per document, there is no upward trend. A peak occurred directly after the introduction of RTX, but it decreased afterward. One possible explanation for this is the diversity in the applications explored in later years that are further apart from the original design of RT cores and specific to a particular problem. 


\begin{table}[]
\caption{Main information on publications.}
\centering
\begin{tabular}{p{0.35\textwidth}l}
\hline
\textbf{Timespan}                  & \textbf{2010:2025} \\ \hline
\textbf{Sources}                   & 49                 \\
\textbf{Documents}                 & 59                 \\
\textbf{Document Average Age}      & 2.97               \\
\textbf{Average citations per doc} & 11.37              \\
\textbf{References}                & 2358               \\
\textbf{Documents Type}            &                    \\
Article                            & 27                 \\
Conference Paper                   & 31                 \\
Review                             & 1                  \\
Document Contents                  &                    \\
Keywords Plus (ID)                 & 586                \\
Author's Keywords (DE)             & 156                \\
\textbf{Authors}                   &                    \\
Authors                            & 173                \\
Authors of single-authored docs    & 3                  \\
\textbf{Authors Collaboration}     &                    \\
Single-authored documents          & 3                  \\
Co-Authors per Document            & 3.98               \\
International co-authorships \%    & 22.03              \\ \hline
\end{tabular}
\label{table:main_information}
\end{table}

\subsection{Keywords Co-occurrence}

\begin{figure*}[ht!]
  \centering
  \includegraphics[width=\textwidth]{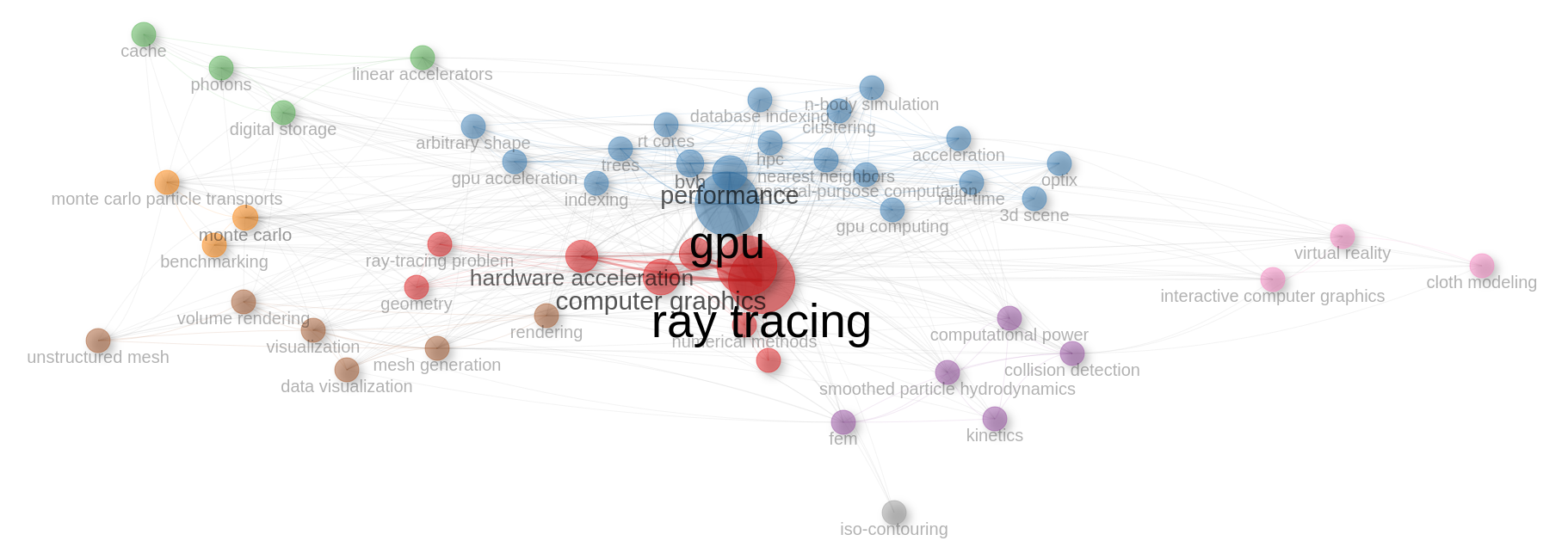}
  \caption{Keywords plus co-occurrence in the set of articles.}
  \label{fig:co-ocurrence}
\end{figure*}

The co-occurrence network in Figure~\ref{fig:co-ocurrence} presents the strongly related keywords in particular clusters, using all keywords, from authors and editors. The most relevant topic, ray tracing, is in its own cluster (red) with keywords related to hardware-accelerated ray tracing and computer graphics. The largest cluster (blue) with the \textit{gpu} keyword refers to the performance of general-purpose computing and includes most several non-graphics tasks of interest for this work. Some general-purpose tasks appear in this cluster (e.g. indexing~\cite{Henneberg2023RTIndeXEH}, nearest neighbors~\cite{zhu2022rtnn} and clustering~\cite{Nagarajan2023RTDBSCANAD}) but most problems are discussed by a single document, so they are not present in the network. The next largest cluster (brown) consists of topics related to visualization; this is in part because the first uses of RT cores beyond ray tracing were related to visualization~\cite{Wald2019RTXBR} which are relevant in scientific simulations, especially geometry and meshes~\cite{Zhao2023RevolutionizingGM}. Similarly, the pink cluster focused on more interactive applications with uses in virtual environments~\cite{Lehericey2013RayTracedCD,IvanovVassilev2021RealtimeVB}. There are two clusters related to physics simulations, the orange includes simulations based on Monte Carlo~\cite{Bahr2021DevelopmentOA}, while the purple consists specifically of hydrodynamics simulated with FEM~\cite{Chan2018ParticlemeshCI}. Finally, the green cluster is a combination between some documents about particle transport~\cite{Lee2024AGM} and articles exploring hardware modifications of RT cores extending them for general-purpose tasks~\cite{Barnes2024ExtendingGR}. Although hardware development is outside the scope of this work, these articles are an alternative to overcome the limitations of current RT cores.

\subsection{Thematic Map}

\begin{figure*}[ht!]
  \centering
  \includegraphics[width=\textwidth]{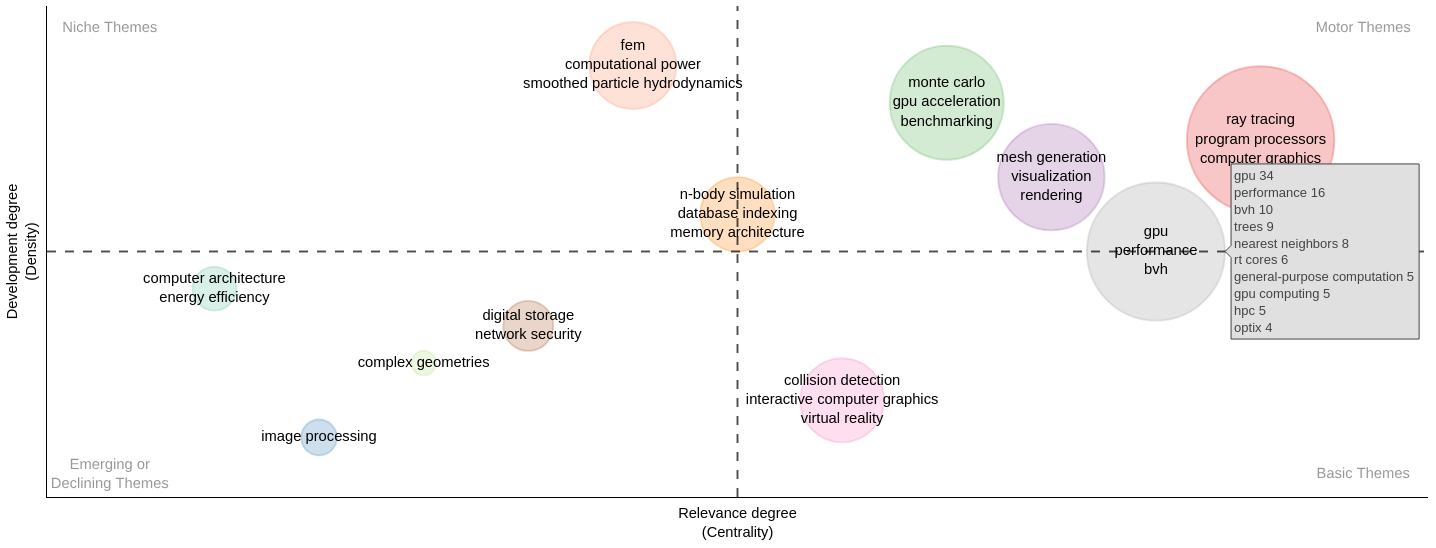}
  \caption{Thematic map}
  \label{fig:thematic-map}
\end{figure*}

The thematic map in Figure~\ref{fig:thematic-map} shows clusters built from all keywords, and their development and relevance among the documents in the bibliometric analysis, i.e. in the context of RT cores. Thus, as expected, ray tracing is a motor theme. Other motor themes include meshes which are used for visualization and rendering~\cite{Morrical2022QuickCA}, and are a key component in several scientific simulations~\cite{Martin2022TheDA}. Monte Carlo simulations are also motor theme in a separate cluster, most documents of this cluster discuss particle transport. The gray cluster, about general-purpose computations is on the edge between being a basic theme and a motor theme. The box with all the keywords in this cluster is included to clarify that it is about general-purpose computing and it includes some non-graphic tasks. Another basic theme includes topics related to interactive applications and virtual environments~\cite{Lehericey2013RayTracedCD,IvanovVassilev2021RealtimeVB}. More specific physics simulation methods like hydrodynamics simulated with FEM are niche themes~\cite{Chan2018ParticlemeshCI}. Some non-graphics tasks that were niche themes are transitioning to motor themes (n-body simulation~\cite{Nagarajan2025RTBarnesHutAB}, database indexing~\cite{Geng2025LibRTSAS}) as the field develops. Emerging/declining themes are specific problems including dexelization~\cite{Inui2020FastDO} and solving PDEs on complex geometries~\cite{Sawhney2023WalkOS} or topics related to hardware development~\cite{Woulfe2017AHF,Barnes2024ExtendingGR}.

\subsection{RT reformulation}

A list of general-purpose problems solved with RT cores is included in Table~\ref{table:speedups}. The reformulation of non-graphics problems to the ray tracing model varies greatly depending on the specific problem. Some problems don't require any reformulation. For example, physics simulation of radiation~\cite{Hashinoki2023ImplementationOR} or particles movement~\cite{Chan2018ParticlemeshCI} can use the same simplified model that ray tracing uses simulating light rays instead of light waves. The only difference is the type of wave simulated, or the particle simulated when thinking of light as photons traveling through space. In this case the only necessary considerations for the ray traced solution is to use data structures compatible with RT cores. Most problems require a novel approach and different perspective. Even some geometric problem that already uses 3D space and the same coordinate system as ray tracing requires a certain level of reformulation. For example, to find the nearest neighbors~\cite{zhu2022rtnn} it is necessary to "reverse" the search radius as explained in the introduction. Other problems need to first map data and queries to 3D space to be compatible with the ray tracing model~\cite{Henneberg2023RTIndeXEH}.

\begin{table*}[htbp]
\centering
\caption{Speedup of RT solutions over state-of-the-art methods}
\begin{filecontents*}{prob.csv}
Problem,Best Speedup,Worst Speedup,Improves?,Avg
Penetration Depth~\cite{Kim2025RTPDPD},5.33,0.93,4,3.2
SpMM~\cite{Zhang2025RTSpMSpMHR},4.02,1.1,5,3.46
BFS~\cite{Xiao2025ACS},1.4,0.4,1,0.54
Triangle Counting~\cite{Xiao2025ACS},2.0,0.4,3,1.07
Set Intersection~\cite{Xiao2025ACS},10.0,0.4,2,1.26
Binary Search~\cite{Xiao2025ACS},2.0,0.5,3,1.15
Point Queries~\cite{Geng2025LibRTSAS},85.1,2.0,5,50.91
Range Queries~\cite{Geng2025LibRTSAS},94.0,1.91,5,55.94
Barnes-Hut~\cite{Nagarajan2025RTBarnesHutAB},93.0,6.0,5,49.41
Discrete CD~\cite{Sui2024HardwareAcceleratedRT},2.8,0.5,4,1.92
Continuous CD~\cite{Sui2024HardwareAcceleratedRT},3.0,0.3,4,1.91
Range Queries~\cite{Henneberg2024MoreBF},1.5,1.0,5,1.44
RMQ~\cite{Meneses2024RTXRMQ},2.3,0.2,2,0.57
Line-Segment Intersection~\cite{Geng2024RayJoinFA},2.4,1.5,5,2.11
Point in Polygon~\cite{Geng2024RayJoinFA},22.2,2.1,5,15.12
Non-euclidean kNN~\cite{Mandarapu2023ArkadeKN},200.0,1.6,5,112.33
ANN~\cite{Liu2023JUNOOH},8.5,2.2,5,6.34
Outlier Detection~\cite{Wang2024RTODEO},9.9,2.0,5,7.34
Index Scan~\cite{Lv2024RTScanES},4.7,1.2,5,3.95
kNN~\cite{Nagarajan2023RTkNNSUU},200.0,2.0,5,109.86
Particle Simulation~\cite{zhao2023leveraging},1.6,1.1,5,1.51
Point Queries~\cite{Henneberg2023RTIndeXEH},7.6,0.2,3,1.7
Radio Wave Propagation~\cite{Hashinoki2023ImplementationOR},5.25,4.86,5,3.8
DBSCAN~\cite{Nagarajan2023RTDBSCANAD},4.0,1.3,4,2.71
Point Location~\cite{Morrical2022AcceleratingUM},6.6,1.7,4,3.95
FRNN~\cite{zhu2022rtnn},10.0,1.0,5,7.94
Particle Tracking~\cite{Wang2022AnGP},2.0,1.08,5,1.85
Graph Drawing~\cite{zellmann2020accelerating},13.0,4.0,5,8.76
Space Skipping~\cite{Morrical2019EfficientSS},7.0,3.0,5,5.16
Segmentation~\cite{Petrescu2019GPUSR},6.2,3.1,5,4.61
Particle-Mesh Coupling~\cite{Chan2018ParticlemeshCI},1.47,0.96,4,1.3
Infrared Radiation~\cite{Liu2025RayTC},13.41,7.38,5,8.47
Particle Transport~\cite{Cui2024RTSRTAM},1.67,1.0,5,1.59
Particle Transport~\cite{Salmon2019ExploitingHR},1.5,1.1,5,1.43
Voxelization~\cite{Schwarz2010FastPS},18.2,2.5,5,12.42
\end{filecontents*}
\csvautotabular{prob.csv}
\label{table:speedups}
\end{table*}

While each problem requires a specific mapping and reformulation, in broad terms they consist of three parts. First, the data needs to be represented as objects in a geometric structure. Second, the queries are mapped to a ray launch, with a starting position, movement vector and magnitude (ray time in ray tracing terminology). Third, the intersection between the ray and an object either produces an operation between them or indicates that the object is the result of the query. The three parts need to be designed coherently with each other such that the ray hits all and only objects relevant to the specific query realized.

In addition, for better performance it is necessary to consider ray tracing strengths and trade-offs. One of its most important characteristics is pruning the tree traversal to avoid unnecessary work~\cite{Morrical2019EfficientSS}, so the geometric structure  should ideally facilitate that. Ray tracing cores are also faster than CUDA cores for tasks with scattered memory access~\cite{Ha2024GeneralizingRT}; when the BVH includes all the necessary data to answer a query, it reduces the number of memory accesses to global memory~\cite{Morrical2022AcceleratingUM}. In the opposite case, however, it can increase memory accesses and cause memory contention between RT cores and CUDA cores~\cite{Zhang2025RTSpMSpMHR}. 

Table~\ref{table:speedups} shows the speedups of these problems recently mentioned compared with standard state-of-the-art solutions. The speedup data was collected from the articles that performed and presented time comparisons between different methods. The reported speedups are with respect to the fastest GPU solution without using RT cores. In the cases where the article only reported time, the speedups were manually calculated. The table presents speedups in the best and worst case scenarios. Average speedups were not considered in part because not only because of the difficulty of obtaining it when not reported, but mostly because determining how representative they are requires a deep understanding of each problem and field. Also, not all authors select test cases the same way, some may choose the most diverse possible tests, others intend to represent the most common use cases. The table also includes the column \textit{Improves?} which consists of a simple scale from 1 to 5 on the amount of cases where the ray tracing solutions performs better. A value of 5 means the RT solution is always or almost always faster, 1 means it is definitely slower and the rest are the middle-ground. To combine the three previous columns an average is calculated by interpreting the scale as the probability of encountering a good case $ p = \frac{\text{Improves? - 0.5}}{5}$. Then the average is calculated as: 
$$\log S_{avg} = p\log S_{best} + (1-p) \log S_{worst} $$
The use of logarithm is necessary since speedup is a multiplicative metric. While the speedups depend on the problems and the range is broad, it is demonstrated that RT cores can greatly increase the performance of some solutions. While kNN~\cite{Nagarajan2023RTkNNSUU} and its non-euclidean variant~\cite{Mandarapu2023ArkadeKN} stand out with speedups of $200\times$ in the best case scenario, most problems have significant speedups even in the worst cases.

\subsection{Problem Characteristics}

The selected problems, in their original formulation (i.e., before the RT core reformulation), were characterized with categorical variables using different taxonomies as shown in Table~\ref{table:problems}. 
Most of the category naming scheme comes from the literature and is familiar to the HPC community. Nevertheless, the difficulty lies in selecting the taxonomies and categories, and organizing them such that they are relevant to RT cores. The only new taxonomy is the RT mapping consisting of the categories direct, native, and abstract. Problems with direct mapping do not require a reformulation to be solved with ray tracing. Native problems use a 3D coordinate system (or one of lower dimensionality) and problems with an abstract mapping require a complete reformulation.

\begin{table*}[ht!]
\caption{Problem characterization.}

\begin{subtable}[h]{\textwidth}
\centering
\footnotesize
\begin{filecontents*}{characteristics1.csv}
Problem;Query Type;Input Type;Dominant Cost;RT Mapping;Domain
Penetration Depth~\cite{Kim2025RTPDPD};Geometry;Mesh;Amortized;Native;Space
SpMM~\cite{Zhang2025RTSpMSpMHR};Indexing;Index;Build;Abstract;Grid
BFS~\cite{Xiao2025ACS};Indexing;Index;Build;Abstract;Graph
Triangle Counting~\cite{Xiao2025ACS};Geometry;Mesh;Build;Native;Graph
Set Intersection~\cite{Xiao2025ACS};Indexing;Index;Query;Abstract;Space
Binary Search~\cite{Xiao2025ACS};Indexing;Index;Query;Abstract;Space
Point Queries~\cite{Geng2025LibRTSAS, Henneberg2023RTIndeXEH};Indexing;Index;Query;Abstract;Index
Range Queries~\cite{Geng2025LibRTSAS, Henneberg2024MoreBF};Indexing;Index;Query;Abstract;Index
Barnes-Hut~\cite{Nagarajan2025RTBarnesHutAB};Simulation;Points;Build;Abstract;Space
Discrete CD~\cite{Sui2024HardwareAcceleratedRT};Geometry;Mesh;Query;Native;Space
Continuous CD~\cite{Sui2024HardwareAcceleratedRT};Geometry;Mesh;Query;Native;Space
RMQ~\cite{Meneses2024RTXRMQ};Indexing;Index;Query;Abstract;Space
Line-Segment Intersection~\cite{Geng2024RayJoinFA};Geometry;Mesh;Query;Abstract;Space
Point in Polygon~\cite{Geng2024RayJoinFA};Geometry;Mesh;Query;Native;Space
Non-euclidean kNN~\cite{Mandarapu2023ArkadeKN};Proximity;Points;Query;Native;Space
ANN~\cite{Liu2023JUNOOH};Proximity;Points;Query;Abstract;Space
Outlier Detection~\cite{Wang2024RTODEO};Proximity;Points;Build;Native;Space
Index Scan~\cite{Lv2024RTScanES};Indexing;Index;Query;Abstract;Index
kNN~\cite{Nagarajan2023RTkNNSUU};Proximity;Points;Query;Abstract;Space
Particle Simulation~\cite{zhao2023leveraging};Simulation;Points;Build;Native;Space
Radio Wave Propagation~\cite{Hashinoki2023ImplementationOR};Simulation;Mesh;Build;Direct;Space
DBSCAN~\cite{Nagarajan2023RTDBSCANAD};Proximity;Points;Amortized;Native;Space
Point Location~\cite{Morrical2022AcceleratingUM};Proximity;Points;Build;Native;Space
FRNN~\cite{zhu2022rtnn};Proximity;Points;Query;Native;Space
Particle Tracking~\cite{Wang2022AnGP};Simulation;Points;Build;Direct;Space
Graph Drawing~\cite{zellmann2020accelerating};Proximity;Points;Build;Native;Graph
Space Skipping~\cite{Morrical2019EfficientSS};Visualization;Mesh;Amortized;Direct;Space
Segmentation~\cite{Petrescu2019GPUSR};Visualization;Grid;Amortized;Native;Grid
Particle-Mesh Coupling~\cite{Chan2018ParticlemeshCI};Simulation;Mesh;Build;Native;Space
Infrared Radiation~\cite{Liu2025RayTC};Simulation;Mesh;Amortized;Direct;Space
Particle Transport~\cite{Lee2024AGM, Cui2024RTSRTAM, Salmon2019ExploitingHR};Simulation;Points;Amortized;Direct;Space
Voxelization~\cite{Schwarz2010FastPS};Visualization;Mesh;Amortized;Native;Space
\end{filecontents*}
\csvautotabular[separator=semicolon]{characteristics1.csv}
\label{table:problems1}
\end{subtable}

\vspace{10px}

\begin{subtable}[h]{\textwidth}
\centering
\footnotesize
\begin{filecontents*}{characteristics2.csv}
Problem;Static/Dynamic;Memory Access;Workload Class;Determinism;Complexity
Penetration Depth~\cite{Kim2025RTPDPD};Static;Hierarchical;Compute-bound;Exact;Quadratic
SpMM~\cite{Zhang2025RTSpMSpMHR};Static;Scattered;Memory-bound;Exact;Linear
BFS~\cite{Xiao2025ACS};Static;Hierarchical;Latency-bound;Exact;Linear
Triangle Counting~\cite{Xiao2025ACS};Static;Scattered;Memory-bound;Exact;Quadratic
Set Intersection~\cite{Xiao2025ACS};Static;Scattered;Latency-bound;Exact;Subquadratic
Binary Search~\cite{Xiao2025ACS};Static;Hierarchical;Compute-bound;Exact;Logarithmic
Point Queries~\cite{Geng2025LibRTSAS, Henneberg2023RTIndeXEH};Static;Scattered;Compute-bound;Exact;Logarithmic
Range Queries~\cite{Geng2025LibRTSAS, Henneberg2024MoreBF};Static;Scattered;Memory-bound;Exact;Logarithmic
Barnes-Hut~\cite{Nagarajan2025RTBarnesHutAB};Dynamic;Hierarchical;Compute-bound;Approximate;Quasilinear
Discrete CD~\cite{Sui2024HardwareAcceleratedRT};Static;Scattered;Compute-bound;Exact;Quasilinear
Continuous CD~\cite{Sui2024HardwareAcceleratedRT};Static;Scattered;Compute-bound;Exact;Quasilinear
RMQ~\cite{Meneses2024RTXRMQ};Static;Scattered;Compute-bound;Exact;Constant
Line-Segment Intersection~\cite{Geng2024RayJoinFA};Static;Scattered;Compute-bound;Exact;Quasilinear
Point in Polygon~\cite{Geng2024RayJoinFA};Static;Scattered;Compute-bound;Exact;Logarithmic
Non-euclidean kNN~\cite{Mandarapu2023ArkadeKN};Static;Scattered;Memory-bound;Exact;Logarithmic
ANN~\cite{Liu2023JUNOOH};Static;Scattered;Memory-bound;Approximate;Logarithmic
Outlier Detection~\cite{Wang2024RTODEO};Static;Scattered;Memory-bound;Approximate;Quasilinear
Index Scan~\cite{Lv2024RTScanES};Static;Scattered;Memory-bound;Exact;Linear
kNN~\cite{Nagarajan2023RTkNNSUU};Static;Scattered;Memory-bound;Exact;Logarithmic
Particle Simulation~\cite{zhao2023leveraging};Dynamic;Hierarchical;Compute-bound;Exact;Quasilinear
Radio Wave Propagation~\cite{Hashinoki2023ImplementationOR};Static;Scattered;Compute-bound;Approximate;Quasilinear
DBSCAN~\cite{Nagarajan2023RTDBSCANAD};Static;Scattered;Memory-bound;Exact;Quasilinear
Point Location~\cite{Morrical2022AcceleratingUM};Static;Hierarchical;Compute-bound;Exact;Logarithmic
FRNN~\cite{zhu2022rtnn};Static;Scattered;Memory-bound;Exact;Logarithmic
Particle Tracking~\cite{Wang2022AnGP};Static;Hierarchical;Memory-bound;Exact;Linear
Graph Drawing~\cite{zellmann2020accelerating};Static;Scattered;Memory-bound;Heuristic;Quasilinear
Space Skipping~\cite{Morrical2019EfficientSS};Static;Scattered;Memory-bound;Heuristic;Logarithmic
Segmentation~\cite{Petrescu2019GPUSR};Static;Scattered;Memory-bound;Heuristic;Linear
Particle-Mesh Coupling~\cite{Chan2018ParticlemeshCI};Dynamic;Scattered;Compute-bound;Approximate;Linear
Infrared Radiation~\cite{Liu2025RayTC};Static;Scattered;Compute-bound;Approximate;Linear
Particle Transport~\cite{Lee2024AGM, Cui2024RTSRTAM, Salmon2019ExploitingHR};Static;Scattered;Compute-bound;Approximate;Linear
Voxelization~\cite{Schwarz2010FastPS};Static;Scattered;Memory-bound;Exact;Linear
\end{filecontents*}
\csvautotabular[separator=semicolon]{characteristics2.csv}
\label{table:problems2}
\end{subtable}
\label{table:problems}

\end{table*}

To understand the relationship between these categories and the speedups, a Kruskal-Wallis test was performed to find taxonomies with statistically significant differences. This test was chosen because it is not parametric, does not rely on the mean, and it is robust to outliers. The result of the test revealed significant differences in three taxonomies: worst case of \textit{Query Type} (p = 0.013),  worst case of  \textit{Workload Class} (p = 0.016), and average of \textit{Determinism} (p = 0.005). Then, a pair-wise test based on the energy distance statistical test was performed. The tests where applied with python using the libraries scipy and hyppo for \textit{Kruskal-Wallis} and \textit{energy-distance} respectively.

\begin{figure*}[ht!]

\begin{subfigure}[h]{0.35\textwidth}
    \includegraphics[scale=0.4]{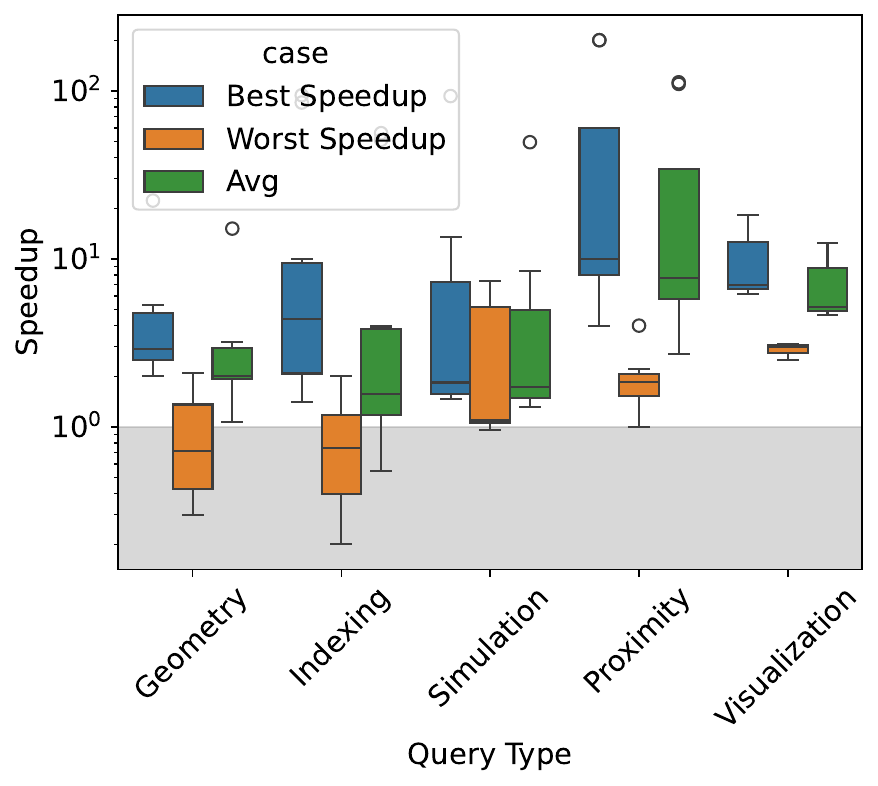}
    \caption{Speedup depending on query type}
    \label{fig:query-type}
\end{subfigure}
\begin{subfigure}[h]{0.35\textwidth}
    \includegraphics[scale=0.4]{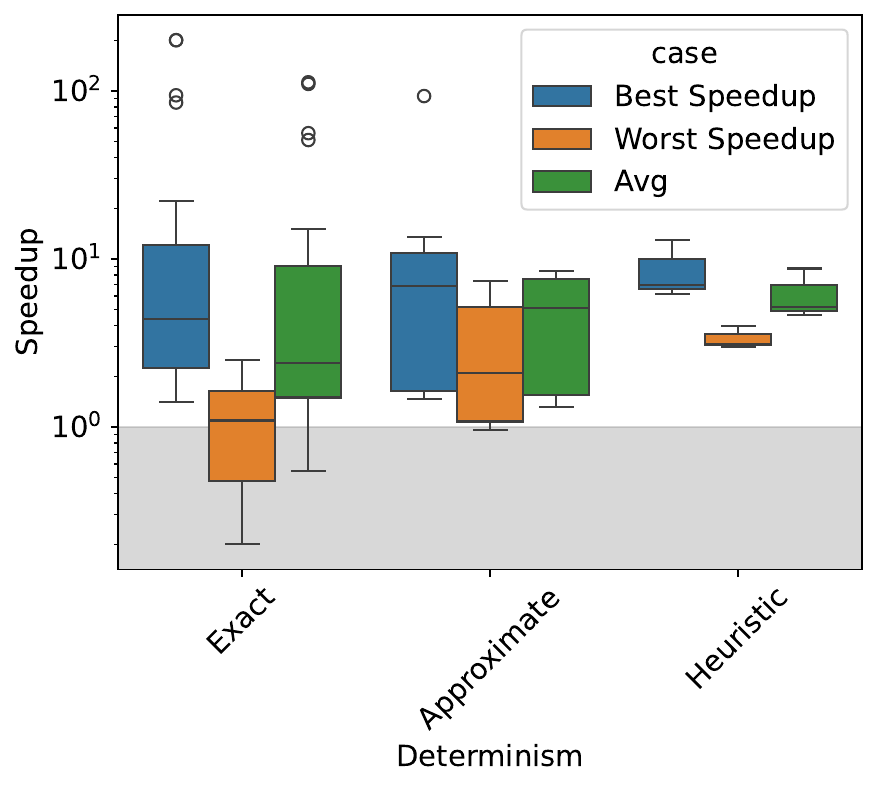}
    \caption{Speedup depending on determinism}
    \label{fig:determinism}
\end{subfigure}
\begin{subfigure}[h]{0.35\textwidth}
    \includegraphics[scale=0.4]{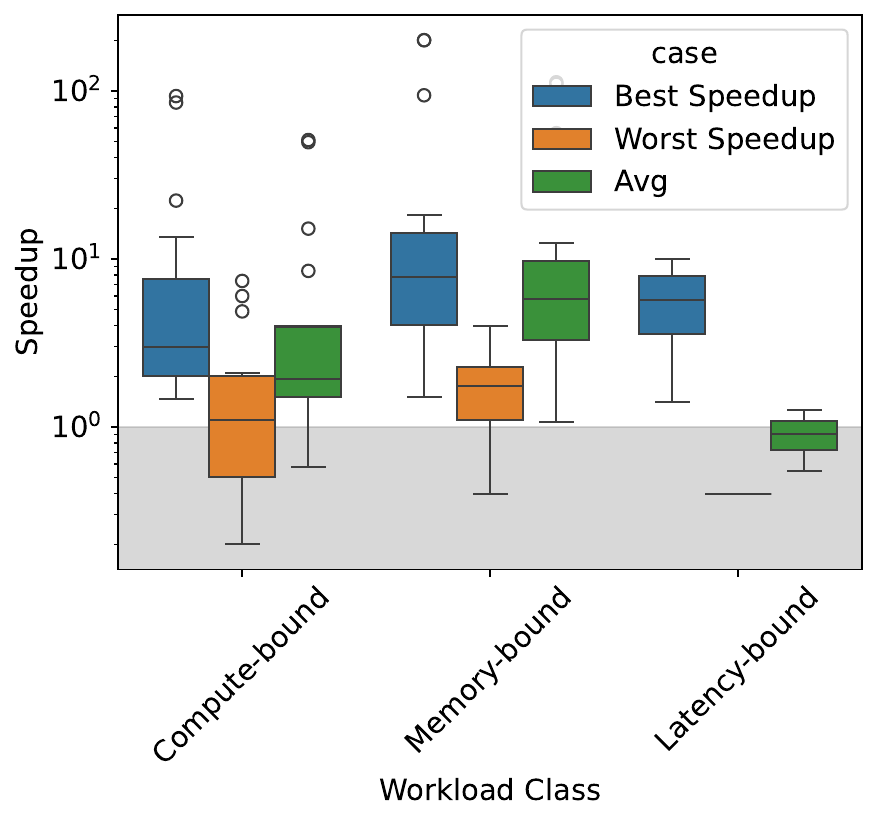}
    \caption{Speedup depending on workload class}
    \label{fig:workload-class}
\end{subfigure}

\caption{Speedup distributions for categories with statistically significant difference}
\label{fig:categories-speedups}
\end{figure*}

Figure~\ref{fig:categories-speedups} shows plots these three taxonomies with boxplots summarizing the distribution of speedups per category. In Figure~\ref{fig:query-type} the speedup according to query type can be appreciated.  Problems usually solved with software implementations of ray tracing such as simulations, or problems similar to it such as visualization queries, never have speedups lower than 1. This demonstrates that  RT cores are always a better option over a software alternative, however, they don't necessarily have the highest speedups. The problems most benefited by RT cores are proximity queries, which consist of nearest neighbors and its variants~\cite{zhu2022rtnn,Mandarapu2023ArkadeKN}. This is a particularly well suited problem for ray tracing because searching with a limited radius and with an infinitesimal ray takes advantage of the BVH pruning and severely limits the possible intersection between rays and spheres. On the other hand, geometric and indexing queries perform worse than the state-of-art solutions in some cases, although there are some outliers with significant performance gains~\cite{Geng2025LibRTSAS}.

For reasons similar to proximity queries, works on heuristics and approximations exhibit speedups over 1 as shown in Figure~\ref{fig:determinism}. They are designed to reduce the amount of work performed matching the pruning of ray tracing~\cite{zellmann2020accelerating}. In the case of aproximations it even allows sacrifices in the accuracy of the result~\cite{Liu2023JUNOOH,Nagarajan2025RTBarnesHutAB}. In the case of latency-bound problems (Figure~\ref{fig:workload-class}, the impact on speedup is unclear as there are only two examples, BFS and set intersection, and there are other important factors to consider as well~\cite{Xiao2025ACS}. In general, reducing memory accesses in BFS does not improve performance because of latency, and even building the BVH takes more time than traversal with CUDA cores. Also, BFS does not take advantage of the biggest strength of RT cores, to avoid work, as all nodes need to be visited, so there is no branches to avoid, nor can traversal be accelerated, because there are operations to perform on internal nodes. Similarly, in set intersections where a lot of ray-intersection checks are needed, the RT methods perform worse than CUDA alternatives, but in the opposite case it can achieve speedups up to $10\times$.

The characteristics in Table~\ref{table:problems} were also used to cluster the problems such that all categories are considered. For this the dimensionality was first reduced with a multiple correspondence analysis (MCA) with 3 components, then the clustering was performed with kmeans. The most important characteristics for each component are the following:


\begin{itemize}
\item \textbf{Component \#0:} Query Type Indexing (14\%), Input Type Index (14\%), RT Mapping abstract (9\%)
\item \textbf{Component \#1:} Query Type Visualization (16\%), Input Type Grid (15\%), Deterministic Class Heuristic (13\%)
\item \textbf{Component \#2:} Query Type Simulation (12\%), Complexity Linear (11\%), Query Type Geometry (10\%)
\end{itemize}

\begin{figure}
  \centering
  \includegraphics[width=0.5\textwidth]{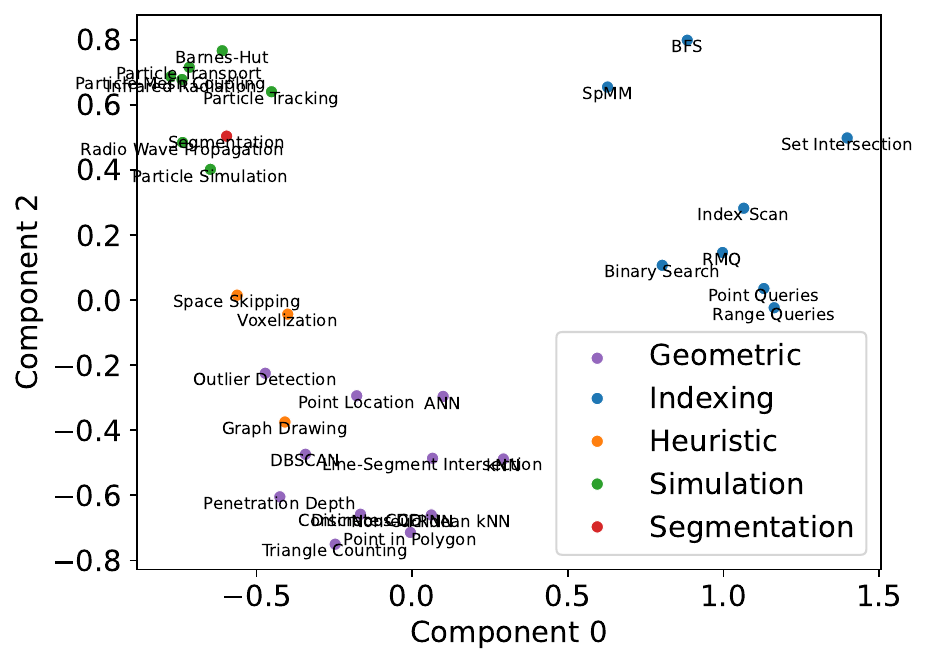}
  \caption{Dimensionality reduction and clusters}
  \label{fig:clusters}
\end{figure}

\begin{figure}
  \centering
  \includegraphics[width=0.5\textwidth]{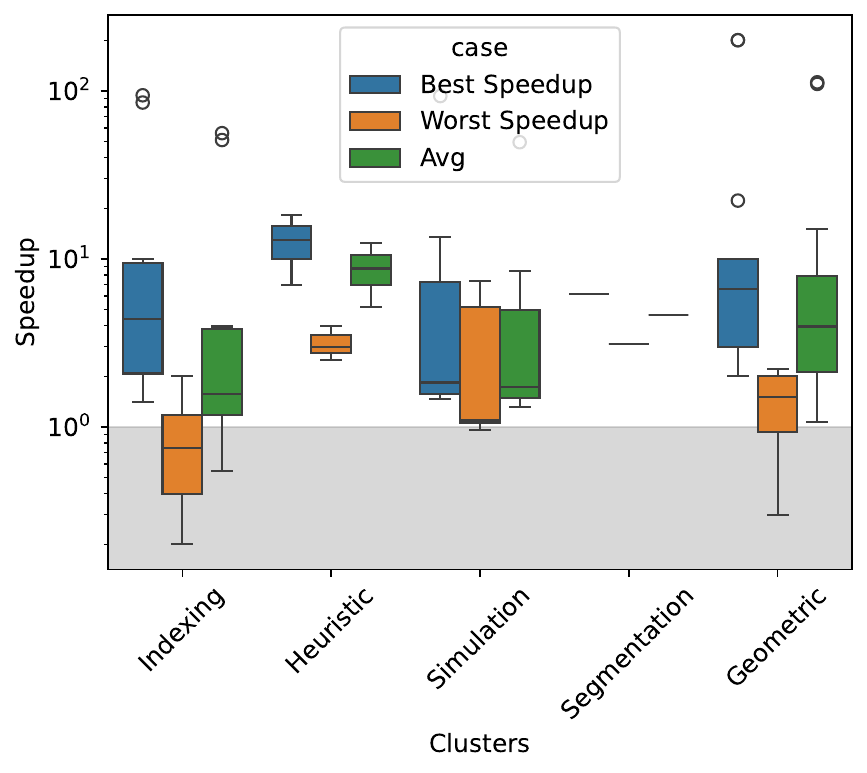}
  \caption{Speedup distribution of clusters}
  \label{fig:clusters_speedup}
\end{figure}

Figure~\ref{fig:clusters} shows components 1 and 2 of the MCA analysis and the resulting 5 clusters. The analysis was repeated with 7 MCA components, sufficient to cover 70\% of the variance, and it resulted in the same 5 clusters. The clusters can be denominated as: indexing, heuristic, simulations, segmentation and geometric, with proximity queries included among geometric problems. Figure~\ref{fig:clusters_speedup} shows the speedups distributions obtained from these clusters. This is a compact summary of all categories, with the exception that proximity problems are not recognized as a separate category. In general, problems closer to ray tracing never perform worse, the impact of heuristics is highlighted and indexing and geometric are to varied to consistently impact speedup. The segmentation problem is a special case as there is only one work as of today~\cite{Petrescu2019GPUSR}; this limits inferring a cluster for this type of problem.

From the results one can start realizing that some problems benefit more from work reduction, rather than the hardware acceleration of ray launches. Several works have taken advantage of that by reducing the size of queries and the amount  of objects close to the path of a ray. To solve kNN~\cite{Nagarajan2023RTkNNSUU}, the closest neighbors are searched by iteratively increasing the search radius and rebuilding the geometric scene each iteration. Thus, in the firsts iterations there a lower amount of intersections because not all neighbors can be found and in the latest iteration the intersections do not increase because rays that already found their neighbor do not need to be launched again. Another example is present in database indexing, where indexing groups of keys instead of keys, reducing the size of the BVH and the memory usage~\cite{Henneberg2024MoreBF}. However, even without reducing the BVH size, a better arrangement of the objects in it can cause a similar effect. A clear example is a work proposed to solve RMQ queries~\cite{Meneses2024RTXRMQ} through small ray tracing queries. This required a more complex and larger geometric scene, but it was arranged such that rays could traverse it faster.

\subsection{RT Cores Limitations}

The greatest limitation of RT cores is its rigid model~\cite{Barnes2024ExtendingGR}. This is due to their focus solely on graphic tasks and the challenges that need to be overcome to solve other problems. Their implementation is so focused that it works as a kind of black box~\cite{Sanzharov2019ExaminationOT}. This is understandable, since it allowed NVIDIA to optimize the implementation for maximum gains on each new GPU. However, it loses possible performance gains in several types of problem, especially those based on tree traversal and hierarchical search~\cite{Ha2024GeneralizingRT,Barnes2024ExtendingGR}. In the current state, the programmer does not have access to the internal nodes of the BVH, so the only way to represent a tree within the ray tracing model is treating all nodes as leafs of the BVH duplicating the amount of nodes in it~\cite{Xiao2025ACS,Nagarajan2025RTBarnesHutAB}.

An important limitation for indexing is the use of memory~\cite{Meneses2024RTXRMQ}. In indexing problems where data consist of single numbers, mapping one number to one triangle requires defining the three corners of the triangle, each with tree coordinates, multiplying by nine the memory usage even without considering the size of the BVH~\cite{Henneberg2023RTIndeXEH}. Also, since these coordinates use only FP32 precision, any problem requiring more precise results need to account for that in the construction of the geometry, and then perform extra computations with more precision and without taking advantage of hardware-accelerated intersection checks~\cite{Geng2024RayJoinFA,Wang2024RTODEO}. The 3D space is also a problem for higher dimensional problems that has not been properly explored. Approximate nearest neighbor is the only problem in which an arbitrary dimension has been mapped to RT cores~\cite{Liu2023JUNOOH}. 

Another disadvantage of RT cores that limits performance is the inherent thread divergence of the ray tracing algorithm. This is not an issue for problems with inherent thread divergence, but it can impact the performance of problems like binary search~\cite{Xiao2025ACS}, even if not coalesced, the memory access in binary search when performing sorted queries has a high cache locality, especially in the first iterations. Another impact in performance of RT cores solutions is the context switching between RT cores and CUDA cores~\cite{Geng2024RayJoinFA}. RT core only perform the tree traversal and intersection checks with triangles, all other operations are performed by CUDA cores, and require data to be transferred. This affects problems that need to compute operations with different objects or can't use triangles. In this case the context is switched after every intersection back and forth between finding more intersections and processing the processing the intersections. The best case are problems that only search the closest hit with triangles, in this case the whole traversal takes place in RT cores and only after the closest hit is found do CUDA core performs the necessary operations.

\section{Conclusions}
This work presented the current state and progress of research on RT cores for non-graphic tasks. In addition, these solutions were evaluated to identify common features, performance gains and limitations of RT cores applied in general-purpose computing. Answering RQ1, most applications are related to physics simulations and on particular geometric or indexing queries, with some focus on neighbor search and database indexing. For RQ2, 32 distinct applications were found, with reformulations to the RT model varying broadly depending on the problem. Especially on problems dissimilar to ray tracing, but all of them share challenges building a geometric scene and mapping queries to ray launches. Regarding RQ3, speedups cover a wide range, some solutions are several times slower in the worst case while others reach up to $200\times$ speedups. A pattern observed is that problems similar to ray tracing exhibit performance gains in all scenarios. Continuing with RQ4, the problems that benefit the most from RT cores are nearest neighbor search, its variants, and problems that can use heuristics to avoid performing unnecessary work. This means that some problems benefit more from work reduction, rather than the hardware acceleration of ray launches. For this reason, reducing the ray size or the number of intersections per ray can improve performance even if it requires more ray launches. This can also be considered a limitation because objects needs to be far from each other or arranged such that they do not provoke intersection checks when there are no intersections because of the bounding boxes of the BVH. Other limitations of RT cores, in regard to RQ5, are the rigidity of the ray tracing model and the focus of the implementation on rendering tasks restricting the generalization to other problems. 

For future work, it is expected for the search of specific problems with different characteristic to continue. There is also an opportunity to apply some of the existent solutions in more contexts. RT cores have improved by approximately $2\times$ each generation in traversing and building the BVH. An underexplored topic is mapping higher dimensional problems to the ray tracing 3D scene; it would be interesting to see more development in this direction. Similarly, more research on hardware modifications to RT core is needed. Although, this was not the focus of this work, it could lead to new approaches to solve problems with RT cores, and possibly lead to hardware improvements being implemented by NVIDIA. 

\section*{Acknowledgements}
This research was supported by the ANID FONDECYT grants \#1221357 and the Patag\'on Supercomputer of Austral University of Chile (FONDEQUIP \#EQM180042).

\bibliographystyle{acm}
\bibliography{sample}

@article{Wald2019RTXBR,
  title={{RTX} beyond ray tracing: exploring the use of hardware ray tracing cores for tet-mesh point location},
  author={Ingo Wald and Will Usher and Nate Morrical and Laura M. Lediaev and Valerio Pascucci},
  journal={Proceedings of the Conference on High-Performance Graphics},
  year={2019}
}

@article{Morrical2022AcceleratingUM,
  title={Accelerating Unstructured Mesh Point Location With RT Cores},
  author={Nate Morrical and Ingo Wald and Will Usher and Valerio Pascucci},
  journal={IEEE Transactions on Visualization and Computer Graphics},
  year={2022},
  volume={28},
  pages={2852-2866}
}

@article{Wang2022AnGP,
  title={An {GPU}-accelerated particle tracking method for Eulerian-Lagrangian simulations using hardware ray tracing cores},
  author={Bin Wang and Ingo Wald and Nate Morrical and Will Usher and Lin Mu and Karsten Thompson and Richard Hughes},
  journal={Comput. Phys. Commun.},
  year={2022},
  volume={271},
  pages={108221}
}

@article{navarro2014survey,
  title={A survey on parallel computing and its applications in data-parallel problems using {GPU} architectures},
  author={Navarro, Cristobal A and Hitschfeld-Kahler, Nancy and Mateu, Luis},
  journal={Communications in Computational Physics},
  volume={15},
  number={2},
  pages={285--329},
  year={2014},
  publisher={Cambridge University Press}
}

@INPROCEEDINGS{zellmann2020accelerating,
  author={Zellmann, Stefan and Weier, Martin and Wald, Ingo},
  booktitle={2020 IEEE Visualization Conference (VIS)}, 
  title={Accelerating Force-Directed Graph Drawing with {RT} Cores}, 
  year={2020},
  volume={},
  number={},
  pages={96-100}
}

@inproceedings{dakkak2019accelerating,
  title={Accelerating reduction and scan using tensor core units},
  author={Dakkak, Abdul and Li, Cheng and Xiong, Jinjun and Gelado, Isaac and Hwu, Wen-mei},
  booktitle={Proceedings of the ACM International Conference on Supercomputing},
  pages={46--57},
  year={2019}
}

@inproceedings{zhu2022rtnn,
  title={{RTNN}: accelerating neighbor search using hardware ray tracing},
  author={Zhu, Yuhao},
  booktitle={Proceedings of the 27th ACM SIGPLAN Symposium on Principles and Practice of Parallel Programming},
  pages={76--89},
  year={2022}
}

@article{zhao2023leveraging,
  title={Leveraging ray tracing cores for particle-based simulations on {GPU}s},
  author={Zhao, Shiwei and Lai, Zhengshou and Zhao, Jidong},
  journal={International Journal for Numerical Methods in Engineering},
  volume={124},
  number={3},
  pages={696--713},
  year={2023},
  publisher={Wiley Online Library}
}

@article{owens2008gpu,
  title={{GPU} computing},
  author={Owens, John D and Houston, Mike and Luebke, David and Green, Simon and Stone, John E and Phillips, James C},
  journal={Proceedings of the IEEE},
  volume={96},
  number={5},
  pages={879--899},
  year={2008},
  publisher={IEEE}
}

@article{Nagarajan2025RTBarnesHutAB,
  title={{RT-BarnesHut}: Accelerating Barnes-Hut Using Ray-Tracing Hardware},
  author={Vani Nagarajan and Rohan Gangaraju and Kirshanthan Sundararajah and Artem Pelenitsyn and Milind Kulkarni},
  journal={Proceedings of the 30th ACM SIGPLAN Annual Symposium on Principles and Practice of Parallel Programming},
  year={2025}
}

@article{Zhang2025RTSpMSpMHR,
  title={{RTSpMSpM}: Harnessing Ray Tracing for Efficient Sparse Matrix Computations},
  author={Hongrui Zhang and Yunan Zhang and Hung-Wei Tseng},
  journal={Proceedings of the 52nd Annual International Symposium on Computer Architecture},
  year={2025}
}

@article{Meneses2024RTXRMQ,
    title = {Accelerating range minimum queries with ray tracing cores},
    journal = {Future Generation Computer Systems},
    volume = {157},
    pages = {98-111},
    year = {2024},
    issn = {0167-739X},
    doi = {https://doi.org/10.1016/j.future.2024.03.040},
    url = {https://www.sciencedirect.com/science/article/pii/S0167739X24001110},
    author = {Enzo Meneses and Cristóbal A. Navarro and Héctor Ferrada and Felipe A. Quezada}
}

@article{Mandarapu2023ArkadeKN,
  title={Arkade: k-Nearest Neighbor Search With Non-Euclidean Distances using GPU Ray Tracing},
  author={Durga Keerthi Mandarapu and Vani Nagarajan and Artem Pelenitsyn and Milind Kulkarni},
  journal={Proceedings of the 38th ACM International Conference on Supercomputing},
  year={2023}
}

@article{Nagarajan2023RTkNNSUU,
  title={RT-kNNS Unbound: Using RT Cores to Accelerate Unrestricted Neighbor Search},
  author={Vani Nagarajan and Durga Keerthi Mandarapu and Milind Kulkarni},
  journal={Proceedings of the 37th International Conference on Supercomputing},
  year={2023}
}

@article{Nagarajan2023RTDBSCANAD,
  title={RT-DBSCAN: Accelerating DBSCAN using Ray Tracing Hardware},
  author={Vani Nagarajan and Milind Kulkarni},
  journal={2023 IEEE International Parallel and Distributed Processing Symposium (IPDPS)},
  year={2023},
  pages={963-973}
}

@article{Lv2024RTScanES,
  title={RTScan: Efficient Scan with Ray Tracing Cores},
  author={Yangming Lv and Kai Zhang and Ziming Wang and Xiaodong Zhang and Rubao Lee and Zhenying He and Yinan Jing and Xiaoyang Sean Wang},
  journal={Proc. VLDB Endow.},
  year={2024},
  volume={17},
  pages={1460-1472}
}

@article{Xiao2025ACS,
  title={A Case Study for Ray Tracing Cores: Performance Insights with Breadth-First Search and Triangle Counting in Graphs},
  author={Zhixiong Xiao and Mengbai Xiao and Yuan Yuan and Dongxiao Yu and Rubao Lee and Xiaodong Zhang},
  journal={Proceedings of the ACM on Measurement and Analysis of Computing Systems},
  year={2025},
  volume={9},
  pages={1 - 25}
}

@article{Theis2017TheEO,
  title={The End of Moore's Law: A New Beginning for Information Technology},
  author={Thomas N. Theis and H.-S. Philip Wong},
  journal={Computing in Science \& Engineering},
  year={2017},
  volume={19},
  pages={41-50}
}

@article{Dally2020DomainspecificHA,
  title={Domain-specific hardware accelerators},
  author={William J. Dally and Yatish Turakhia and Song Han},
  journal={Communications of the ACM},
  year={2020},
  volume={63},
  pages={48 - 57}
}

@article{Aila2009UnderstandingTE,
  title={Understanding the efficiency of ray traversal on GPUs},
  author={Timo Aila and Samuli Laine},
  journal={Proceedings of the Conference on High Performance Graphics 2009},
  year={2009}
}

@article{Boukaram2019HierarchicalMO,
  title={Hierarchical Matrix Operations on GPUs},
  author={Wajih Halim Boukaram and George M. Turkiyyah and David E. Keyes},
  journal={ACM Transactions on Mathematical Software (TOMS)},
  year={2019},
  volume={45},
  pages={1 - 28}
}

@article{Dally2021EvolutionOT,
  title={Evolution of the Graphics Processing Unit (GPU)},
  author={William J. Dally and Stephen W. Keckler and David Blair Kirk},
  journal={IEEE Micro},
  year={2021},
  volume={41},
  pages={42-51}
}

@article{Navarro2024CATCA,
  title={CAT: Cellular Automata on Tensor Cores},
  author={Crist{\'o}bal A. Navarro and Felipe A. Quezada and Enzo Meneses and H'ector Ferrada and Nancy Hitschfeld},
  journal={IEEE Transactions on Parallel and Distributed Systems},
  year={2024},
  volume={36},
  pages={341-355}
}

@article{Sanzharov2020SurveyON,
  title={Survey of Nvidia RTX Technology},
  author={Vadim Vladimirovich Sanzharov and Vladimir A. Frolov and Vladimir Alexandrovich Galaktionov},
  journal={Programming and Computer Software},
  year={2020},
  volume={46},
  pages={297 - 304}
}

@article{Sarton2023StateoftheartIL,
  title={State‐of‐the‐art in Large‐Scale Volume Visualization Beyond Structured Data},
  author={Jonathan Sarton and Stefan Zellmann and Serkan Demirci and Uğur G{\"u}d{\"u}kbay and W. Alexandre‐Barff and Laurent Lucas and Jean-Michel Dischler and Stefan Wesner and Ingo Wald},
  journal={Computer Graphics Forum},
  year={2023},
  volume={42}
}

@article{Echeverra2024HarnessingTP,
  title={Harnessing the Power of Ray Tracing for Enhanced 3D Map-based Localization using NanoVDB},
  author={Andrea Maybell Pe{\~n}a Echeverr{\'i}a and Paul Kemppi},
  journal={2024 10th International Conference on Mechatronics and Robotics Engineering (ICMRE)},
  year={2024},
  pages={240-246}
}

@article{Meister2021ASO,
  title={A Survey on Bounding Volume Hierarchies for Ray Tracing},
  author={Daniel Meister and Shinji Ogaki and Carsten Benthin and Michael J. Doyle and Michael Guthe and Jiř{\'i} Bittner},
  journal={Computer Graphics Forum},
  year={2021},
  volume={40}
}

@article{Fedus2021SwitchTS,
  title={Switch Transformers: Scaling to Trillion Parameter Models with Simple and Efficient Sparsity},
  author={William Fedus and Barret Zoph and Noam M. Shazeer},
  journal={ArXiv},
  year={2021},
  volume={abs/2101.03961}
}

@article{Narayanan2021EfficientLL,
  title={Efficient Large-Scale Language Model Training on GPU Clusters Using Megatron-LM},
  author={Deepak Narayanan and Mohammad Shoeybi and Jared Casper and Patrick LeGresley and Mostofa Patwary and Vijay Anand Korthikanti and Dmitri Vainbrand and Prethvi Kashinkunti and Julie Bernauer and Bryan Catanzaro and Amar Phanishayee and Matei A. Zaharia},
  journal={SC21: International Conference for High Performance Computing, Networking, Storage and Analysis},
  year={2021},
  pages={1-14}
}

@article{Maksimova2021AbacusSummitAM,
  title={AbacusSummit: A Massive Set of High-Accuracy, High-Resolution N-Body Simulations},
  author={Nina A Maksimova and Lehman H. Garrison and Daniel J. Eisenstein and Boryana Hadzhiyska and Sownak Bose and T.P. Satterthwaite},
  journal={Monthly Notices of the Royal Astronomical Society},
  year={2021}
}

@article{Schade2022BreakingTE,
  title={Breaking the exascale barrier for the electronic structure problem in ab-initio molecular dynamics},
  author={Robert Schade and Tobias Kenter and Hossam Elgabarty and Michael Lass and Thomas D. K{\"u}hne and Christian Plessl},
  journal={The International Journal of High Performance Computing Applications},
  year={2022},
  volume={37},
  pages={530 - 538}
}

@article{CocanaFernandez2019EcoEfficientRM,
  title={Eco-Efficient Resource Management in HPC Clusters through Computer Intelligence Techniques},
  author={Alberto Coca{\~n}a-Fern{\'a}ndez and Emilio San Jos{\'e} Guiote and Luciano S{\'a}nchez and Jos{\'e} Ranilla},
  journal={Energies},
  year={2019}
}

@article{Li2023AnalyzingRU,
  title={Analyzing Resource Utilization in an HPC System: A Case Study of NERSC Perlmutter},
  author={Jie Li and George Michelogiannakis and Brandon Cook and Dulanya Cooray and Yong Chen},
  journal={ArXiv},
  year={2023},
  volume={abs/2301.05145}
}

@article{Gerber2011LargeSC,
  title={Large Scale Computing and Storage Requirements for High Energy Physics},
  author={Richard A. Gerber and Harvey J. Wasserman},
  journal={Lawrence Berkeley National Laboratory},
  year={2011}
}

@inproceedings{Arima2024OnTC,
  title={On the Convergence of Malleability and the HPC PowerStack: Exploiting Dynamism in Over-Provisioned and Power-Constrained HPC Systems},
  author={Eishi Arima and Isa{\'i}as Compr{\'e}s and Martin Schulz},
  booktitle={ISC Workshops},
  year={2024}
}

@inproceedings{Martin2014PostDennardSA,
  title={Post-Dennard Scaling and the final Years of Moore ’ s Law Consequences for the Evolution of Multicore-Architectures},
  author={Christian Märtin},
  year={2014}
}

@inproceedings{Sueur2010DynamicVA,
  title={Dynamic voltage and frequency scaling: the laws of diminishing returns},
  author={Etienne Le Sueur and Gernot Heiser},
  year={2010}
}

@article{Markidis2018NVIDIATC,
  title={NVIDIA Tensor Core Programmability, Performance \& Precision},
  author={Stefano Markidis and Steven W. D. Chien and Erwin Laure and Ivy Bo Peng and Jeffrey S. Vetter},
  journal={2018 IEEE International Parallel and Distributed Processing Symposium Workshops (IPDPSW)},
  year={2018},
  pages={522-531}
}

@article{Lopez2023MixedPL,
  title={Mixed precision LU factorization on GPU tensor cores: reducing data movement and memory footprint},
  author={Florent Lopez and Th{\'e}o Mary},
  journal={The International Journal of High Performance Computing Applications},
  year={2023},
  volume={37},
  pages={165 - 179}
}

@article{Henneberg2023RTIndeXEH,
  title={RTIndeX: Exploiting Hardware-Accelerated GPU Raytracing for Database Indexing},
  author={Justus Henneberg and Felix Martin Schuhknecht},
  journal={Proc. VLDB Endow.},
  year={2023},
  volume={16},
  pages={4268-4281}
}

@article{Grant2009ATO,
  title={A typology of reviews: an analysis of 14 review types and associated methodologies.},
  author={Maria J. Grant and Andrew Booth},
  journal={Health information and libraries journal},
  year={2009},
  volume={26 2},
  pages={91-108}
}

@article{Budgen2006PerformingSL,
  title={Performing systematic literature reviews in software engineering},
  author={David Budgen and Pearl Brereton},
  journal={Proceedings of the 28th international conference on Software engineering},
  year={2006}
}

@article{Chigbu2023TheSO,
  title={The Science of Literature Reviews: Searching, Identifying, Selecting, and Synthesising},
  author={Uchendu Eugene Chigbu and Sulaiman Olusegun Atiku and Cherley C. Du Plessis},
  journal={Publ.},
  year={2023},
  volume={11},
  pages={2}
}

@article{Liberati2009ThePS,
  title={The PRISMA Statement for Reporting Systematic Reviews and Meta-Analyses of Studies That Evaluate Health Care Interventions: Explanation and Elaboration},
  author={Alessandro Liberati and Douglas G. Altman and Jennifer Marie Tetzlaff and Cynthia D. Mulrow and Peter Christian G{\o}tzsche and John P. A. Ioannidis and Mike Clarke and Philip J. Devereaux and Jos Kleijnen and David Moher},
  journal={PLoS Medicine},
  year={2009},
  volume={6}
}

@article{Page2020TheP2,
  title={The PRISMA 2020 statement: an updated guideline for reporting systematic reviews},
  author={Matthew J. Page and Joanne E. McKenzie and Patrick M. M. Bossuyt and Isabelle Boutron and Tammy C. Hoffmann and Cynthia D. Mulrow and Larissa Shamseer and Jennifer Marie Tetzlaff and Elie A. Akl and Sue E. Brennan and Roger Chou and Julie May Glanville and Jeremy M. Grimshaw and Asbj{\o}rn Hr{\~o}bjartsson and Manoj Mathew Lalu and Tianjing Li and Elizabeth W. Loder and Evan Mayo-Wilson and Steve McDonald and Luke A. McGuinness and Lesley A Stewart and James Thomas and Andrea C. Tricco and Vivian A Welch and Penny F. Whiting and David Moher},
  journal={Systematic Reviews},
  year={2020},
  volume={10}
}

@article{Dongarra2024TheCO,
  title={The co-evolution of computational physics and high-performance computing},
  author={Jack J. Dongarra and David E. Keyes},
  journal={Nature Reviews Physics},
  year={2024},
  volume={6},
  pages={621 - 627}
}

@article{Koch2023HPCIT,
  title={HPC+ in the medical field: Overview and current examples},
  author={Miriam Koch and Claudio Arlandini and Gregory Antonopoulos and Alessia Baretta and Pierre Beaujean and Geert Jan Bex and Marco Evangelos Biancolini and Simona Celi and Emiliano Costa and Lukas Drescher and Vasileios Eleftheriadis and Nur A. Fadel and Andreas Fink and Federica Galbiati and Ilias Hatzakis and Georgios Hompis and Natalie Lewandowski and Antonio Memmolo and Carl Mensch and Dominik Obrist and Valentina Paneta and Panagiotis G. Papadimitroulas and Konstantinos Petropoulos and Stefano Porziani and Georgios Savvidis and Khyati Sethia and Petr Strako\v{s} and Petra Svobodova and Emanuele Vignali},
  journal={Technology and Health Care},
  year={2023},
  volume={31},
  pages={1509 - 1523}
}

@article{Kim2025RTPDPD,
  title={RTPD: penetration depth calculation using hardware-accelerated ray-tracing},
  author={YoungWoo Kim and Sungmin Kwon and Duksu Kim},
  journal={The Visual Computer},
  year={2025},
  volume={41},
  pages={9885 - 9899}
}

@article{Geng2025LibRTSAS,
  title={LibRTS: A Spatial Indexing Library by Ray Tracing},
  author={Liang Geng and Rubao Lee and Xiaodong Zhang},
  journal={Proceedings of the 30th ACM SIGPLAN Annual Symposium on Principles and Practice of Parallel Programming},
  year={2025}
}

@article{Sui2024HardwareAcceleratedRT,
  title={Hardware-Accelerated Ray Tracing for Discrete and Continuous Collision Detection on GPUs},
  author={Sizhe Sui and Luis Sentis and Andrew Bylard},
  journal={2025 IEEE International Conference on Robotics and Automation (ICRA)},
  year={2024},
  pages={16133-16139}
}

@article{Henneberg2024MoreBF,
  title={More Bang for Your Buck(et): Fast and Space-Efficient Hardware-Accelerated Coarse-Granular Indexing on GPUs},
  author={Justus Henneberg and Felix Martin Schuhknecht and Rosina F. Kharal and Trevor Brown},
  journal={2025 IEEE 41st International Conference on Data Engineering (ICDE)},
  year={2024},
  pages={1320-1333}
}

@article{Geng2024RayJoinFA,
  title={RayJoin: Fast and Precise Spatial Join},
  author={Liang Geng and Rubao Lee and Xiaodong Zhang},
  journal={Proceedings of the 38th ACM International Conference on Supercomputing},
  year={2024}
}

@article{Liu2023JUNOOH,
  title={JUNO: Optimizing High-Dimensional Approximate Nearest Neighbour Search with Sparsity-Aware Algorithm and Ray-Tracing Core Mapping},
  author={Zihan Liu and Wentao Ni and Jingwen Leng and Yu Feng and Cong Guo and Quan Chen and Chao Li and Minyi Guo and Yuhao Zhu},
  journal={Proceedings of the 29th ACM International Conference on Architectural Support for Programming Languages and Operating Systems, Volume 2},
  year={2023}
}

@article{Wang2024RTODEO,
  title={RTOD: Efficient Outlier Detection With Ray Tracing Cores},
  author={Ziming Wang and Kaiyin Zhang and Yangming Lv and Yinglong Wang and Zhigang Zhao and Zhenying He and Yinan Jing and Xiaoyang Sean Wang},
  journal={IEEE Transactions on Knowledge and Data Engineering},
  year={2024},
  volume={36},
  pages={9192-9204}
}

@article{Hashinoki2023ImplementationOR,
  title={Implementation of Radio Wave Propagation using RT Cores and Consideration of Programming Models},
  author={Shinya Hashinoki and Satoshi Ohshima and Takahiro Katagiri and Toru Nagai and Tetsuya Hoshino},
  journal={2023 IEEE International Parallel and Distributed Processing Symposium Workshops (IPDPSW)},
  year={2023},
  pages={673-681}
}

@article{Morrical2019EfficientSS,
  title={Efficient Space Skipping and Adaptive Sampling of Unstructured Volumes Using Hardware Accelerated Ray Tracing},
  author={Nate Morrical and Will Usher and Ingo Wald and Valerio Pascucci},
  journal={2019 IEEE Visualization Conference (VIS)},
  year={2019},
  pages={256-260}
}

@article{Petrescu2019GPUSR,
  title={GPU Sparse Ray-Traced Segmentation},
  author={Lucian Petrescu and Anca Morar and Florica Moldoveanu and Alin Dragos Bogdan Moldoveanu},
  journal={IEEE Access},
  year={2019},
  volume={7},
  pages={68511-68521},
  url={https://api.semanticscholar.org/CorpusID:174818776}
}

@article{Chan2018ParticlemeshCI,
  title={Particle–mesh coupling in the interaction of fluid and deformable bodies with screen space refraction rendering},
  author={Ka‐Hou Chan and W. Ke and Sio Kei Im},
  journal={Computer Animation and Virtual Worlds},
  year={2018},
  volume={29}
}

@article{Liu2025RayTC,
  title={Ray tracing core accelerated high performance implementation for infrared radiation signature simulation using reverse Monte Carlo method},
  author={Jiaan Liu and Qitai Eri and Bo Kong and Jifeng Huang and Yue Zhou},
  journal={International Journal of Thermal Sciences},
  year={2025}
}

@article{Lee2024AGM,
  title={A GPU-accelerated Monte Carlo code, RT2 for coupled transport of photon, electron/positron, and neutron},
  author={Chang-Min Lee and Sung-Joon Ye},
  journal={Physics in Medicine \& Biology},
  year={2024},
  volume={69}
}

@article{Cui2024RTSRTAM,
  title={RTSRT: Accelerating Monte Carlo Particle Transport with Ray Tracing Shared Cache Architecture},
  author={Cunhao Cui and Tiejun Li and Jianmin Zhang and Hanqing Li and Changsong Jin and Ruixuan Ren},
  journal={2024 IEEE International Symposium on Parallel and Distributed Processing with Applications (ISPA)},
  year={2024},
  pages={600-606}
}

@article{Salmon2019ExploitingHR,
  title={Exploiting Hardware-Accelerated Ray Tracing for Monte Carlo Particle Transport with OpenMC},
  author={Justin Salmon and Simon McIntosh-Smith},
  journal={2019 IEEE/ACM Performance Modeling, Benchmarking and Simulation of High Performance Computer Systems (PMBS)},
  year={2019},
  pages={19-29}
}

@article{Schwarz2010FastPS,
  title={Fast parallel surface and solid voxelization on GPUs},
  author={Michael Schwarz and Hans-Peter Seidel},
  journal={ACM SIGGRAPH Asia 2010 papers},
  year={2010}
}

@article{Suarez2025EnergyET,
  title={Energy Efficiency trends in HPC: what high-energy and astrophysicists need to know},
  author={Estela Suarez and Jorge Amaya and Martin Frank and Oliver Freyermuth and Maria Girone and Bartosz Kostrzewa and Susanne Pfalzner},
  journal={ArXiv},
  year={2025},
  volume={abs/2503.17283}
}

@inproceedings{Borner2012NetworkST,
  title={Network Science: Theory, Tools, and Practice},
  author={Katy B{\"o}rner and Soma Sanyal and Alessandro Vespignani},
  year={2012}
}

@article{Aria2017bibliometrixAR,
  title={bibliometrix: An R-tool for comprehensive science mapping analysis},
  author={Massimo Aria and Corrado Cuccurullo},
  journal={J. Informetrics},
  year={2017},
  volume={11},
  pages={959-975}
}

@article{Ha2024GeneralizingRT,
  title={Generalizing Ray Tracing Accelerators for Tree Traversals on GPUs},
  author={Dongho Ha and Lufei Liu and Yuan-Hsi Chou and Seokjin Go and Won Woo Ro and Hung-Wei Tseng and Tor M. Aamodt},
  journal={2024 57th IEEE/ACM International Symposium on Microarchitecture (MICRO)},
  year={2024},
  pages={1041-1057}
}

@article{Barnes2024ExtendingGR,
  title={Extending GPU Ray-Tracing Units for Hierarchical Search Acceleration},
  author={Aaron Barnes and Fangjia Shen and Timothy G. Rogers},
  journal={2024 57th IEEE/ACM International Symposium on Microarchitecture (MICRO)},
  year={2024},
  pages={1027-1040}
}

@article{Buscher2024ACS,
  title={A Comprehensive Survey of Isocontouring Methods: Applications, Limitations and Perspectives},
  author={Keno Jann B{\"u}scher and Jan Philipp Degel and Jan Oellerich},
  journal={Algorithms},
  year={2024},
  volume={17},
  pages={83}
}

@article{Zhao2023RevolutionizingGM,
  title={Revolutionizing granular matter simulations by high-performance ray tracing discrete element method for arbitrarily-shaped particles},
  author={Shiwei Zhao and Jidong Zhao},
  journal={Computer Methods in Applied Mechanics and Engineering},
  year={2023}
}

@inproceedings{Lehericey2013RayTracedCD,
  title={Ray-Traced Collision Detection: Interpenetration Control and Multi-GPU Performance},
  author={François Lehericey and Val{\'e}rie Gouranton and Bruno Arnaldi},
  booktitle={EGVE/EuroVR},
  year={2013}
}

@article{IvanovVassilev2021RealtimeVB,
  title={Real-time Velocity Based Cloth Simulation with Ray-tracing Collision Detection on the Graphics Processor},
  author={Tzvetomir Ivanov Vassilev},
  journal={2021 International Conference on Information Technologies (InfoTech)},
  year={2021},
  pages={1-5}
}

@article{Bahr2021DevelopmentOA,
  title={Development of a hardware-accelerated simulation kernel for ultra-high vacuum with Nvidia RTX GPUs},
  author={Pascal R B{\"a}hr and Bruno Lang and Peer Ueberholz and M. Ady and Roberto Kersevan},
  journal={The International Journal of High Performance Computing Applications},
  year={2021},
  volume={36},
  pages={141 - 152}
}

@article{Morrical2022QuickCA,
  title={Quick Clusters: A GPU-Parallel Partitioning for Efficient Path Tracing of Unstructured Volumetric Grids},
  author={Nate Morrical and Alper Sahistan and Uğur G{\"u}d{\"u}kbay and Ingo Wald and Valerio Pascucci},
  journal={IEEE Transactions on Visualization and Computer Graphics},
  year={2022},
  volume={29},
  pages={537-547}
}

@article{Martin2022TheDA,
  title={The Design and Implementation of a Ray-tracing Algorithm for Signal-level Pulsed Radar Simulation Using the NVIDIA{\textregistered} OptiXTM Engine},
  author={Mogamat Yaaseen Martin and Simon Lucas Winberg and Mohammed Yunus Abdul Gaffar and Dave MacLeod},
  journal={J. Commun.},
  year={2022},
  volume={17},
  pages={761-768}
}

@article{Sawhney2023WalkOS,
  title={Walk on Stars: A Grid-Free Monte Carlo Method for PDEs with Neumann Boundary Conditions},
  author={Rohan Sawhney and Bailey Miller and Ioannis Gkioulekas and Keenan Crane},
  journal={ACM Transactions on Graphics (TOG)},
  year={2023},
  volume={42},
  pages={1 - 20}
}

@article{Inui2020FastDO,
  title={Fast Dexelization of Polyhedral Models Using Ray-Tracing Cores of GPU},
  author={Masatomo Inui and Kohei Kaba and Nobuyuki Umezu},
  journal={Computer-Aided Design and Applications},
  year={2020}
}

@article{Woulfe2017AHF,
  title={A hybrid fixed-function and microprocessor solution for high-throughput broad-phase collision detection},
  author={Muiris Woulfe and Michael Manzke},
  journal={EURASIP Journal on Embedded Systems},
  year={2017},
  volume={2017},
  pages={1-15}
}

@article{Sanzharov2019ExaminationOT,
  title={Examination of the Nvidia RTX},
  author={Vadim Vladimirovich Sanzharov and Alexey Gorbonosov and Vladimir Alexandrovich Frolov and Alexey Gennadievich Voloboy},
  journal={GraphiCon'2019 Proceedings. Volume 2},
  year={2019}
}

@misc{NVIDIA_Blackwell_whitepaper, 
    title={{NVIDIA} {RTX} Blackwell {GPU} architecture}, 
    url={https://images.nvidia.com/aem-dam/Solutions/geforce/blackwell/nvidia-rtx-blackwell-gpu-architecture.pdf}, 
    author={NVIDIA Corporation}
}

\end{document}